\documentclass[10pt,english]{article}
\usepackage{babel}
\usepackage{shortvrb}
\usepackage[latin1]{inputenc}
\usepackage{graphicx}
\usepackage{color}
\usepackage{multirow}
\usepackage[colorlinks]{hyperref}
\usepackage[a4paper,margin=2cm]{geometry}
\usepackage{palatino}
\title{Modernizing the ESRF beamline software architecture with generic Python modules}
\hypersetup{
pdftitle={Modernizing the ESRF beamline software architecture with generic Python modules}
}
\begin{document}
\maketitle

%___________________________________________________________________________

\hypertarget{introduction}{}
\section*{Introduction}
\pdfbookmark[0]{Introduction}{introduction}

This article describes the new application software architecture on the 
ESRF beamlines. It will show its basic elements and comment on the level 
of their advancement.

The work has been mainly done in the BLISS group (Beamline Instrument 
Software Support) and by CSS (Certified Scientific Software) for the Spec 
Server. The list of contributers to this project include G. Berruyer,
A. Gobbo, L. Claustre,T. Jouve, J. Klora, E. Papillon, N. Pascal, V. Rey, 
A. Sole, D. Spruce, and G. Swislow.

%___________________________________________________________________________

\hypertarget{the-global-beamline-control-software-structure}{}
\section*{The global beamline control software structure}
\pdfbookmark[0]{The global beamline control software structure}{the-global-beamline-control-software-structure}

The beamline software architecture at the ESRF is built around the commercial
data acquisition package SPEC. SPEC talks to our hardware mainly via the ESRF
TACO device servers. The device servers can either be specific to one type of
device or just represent the hardware channel to talk to a device (i.e. a
serial line). For a small number of device types (i.e stepper motors) which
represent however the majority of hardware items used on the beamlines (i.e. we
have about 3000 motors on our beamline) the code to talk to these device has
been added to SPECs internal codebase. For the rest of devices the access is
done via SPECs macro language. The end user does normally not see a difference
between these methods.

\includegraphics{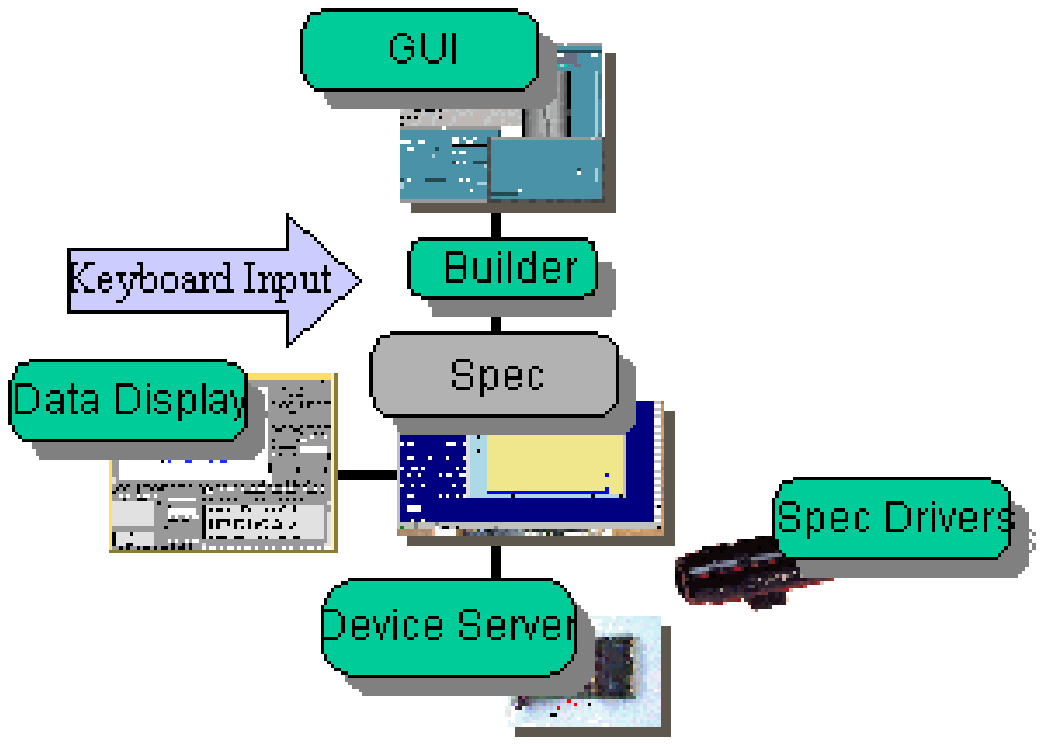}

The data display routines in SPEC are not sufficient for our beamlines. We
have therefore decided to add external data visualization tools (i.e. for
CCD cameras). The communication between the different programs is done 
using shared memory with specific header informations.

GUI user interface are put on top of SPEC. In the past we were using a 
program which mixed the standard user input from the keyboard with the 
input from the GUIs. SPEC sent information back to this program via
a dedicated pipe. Multiple GUIs can be used at the same time. The 
information sent has been distributed to the GUIs in some type of event.

Two types of such GUIs have been used at the ESRF. There is a general SPEC GUI 
written in Motif which had a built in editor for graphical panels. The 
relatively simple panels were used to gather information to start SPEC
macros. No programming had to be done in the GUI. There exists also a
GUI written in Tcl/Tk dedicated to Protein crystallography experiments.
This GUI contains much more dedicated logic but will also in the end send
commands to SPEC to carry out the experiments. It will however do many things
in addition (i.e. add information to a central data base).

%___________________________________________________________________________

\hypertarget{the-new-building-blocks}{}
\section*{The new building blocks}
\pdfbookmark[0]{The new building blocks}{the-new-building-blocks}

The modernization of the control system keeps the general structure 
described in the last paragraph. Individual components are replaced with 
modules and applications written in Python. We can therefore distinguish
the following areas:
\begin{itemize}
\item 
The external program which mixes keyboard input and the GUI communication 
to SPEC has been replaced by a new server mode in SPEC. When started in 
server mode, SPEC listens to requests from clients and sends events 
to them. We have a Python client library which provides access to SPEC with 
an object oriented interface.

\item 
A plotting framework (PyDVT) has been created. The framework defines classes
in the following areas:
\begin{quote}
\begin{itemize}
\item 
visualization (like 2D images or contour plots)

\item 
data sources (like input from a file or from SPECs shared memory)

\item 
user selections (like cuts)

\item 
filters (like fitting routines or color selectors)

\end{itemize}
\end{quote}

It makes it then possible to define the data flow between them.

\item 
Applications for visualization have been created from this framework.
They work either standalone with data from files or integrated in the 
online data visualization with data from SPEC shared memory. Examples of 
these applications are given later.

\item 
A new configuration editor has been created. This editor can be freely 
configured. It integrates the functionality of SPECs config editor and
our own in-house configuration tools.

\item 
A general GUI framework application has been started in Python. 
The application allows to load Python classes and integrate them
into the running system. There is an built in editor to create simpler
panels interactively without any programming.

\end{itemize}

%___________________________________________________________________________

\hypertarget{why-python}{}
\section*{Why Python}
\pdfbookmark[0]{Why Python}{why-python}

Our choice for Python was relatively simple to make. We wanted a scripting
language which can be easily used by the scientists on the beamline. 
Python can be extended with modules in Python or other languages (i.e. C). 
The language is very compact, easy to read, and offers an enormous amount
of builtin functionality. It is largely platform independent. You can read more
on www.python.org.

Unluckily the choice of the toolkit for Python was less obvious. We started 
using the most widely used toolkit Tk (coming from Tcl/Tk). We did have to
program many standard widgets for our projects and the final programs didn't 
have the "look and feel" we wanted. We decided therefore to try Qt with 
Python bindings (www.trolltech.com) and are very happy with the functionality. 
Python/Qt is however not very widely used. We tried therefore to be as 
independent of the graphical toolkit as possible in the basic classes.

%___________________________________________________________________________

\hypertarget{the-spec-server}{}
\section*{The Spec Server}
\pdfbookmark[0]{The Spec Server}{the-spec-server}

SPEC when started in server mode accepts connections on a
socket. Multiple clients can connect and send messages. Clients can
execute commands, talk to motor or counter objects in the server, read
variables (i.e. global variables or motor parameters) from the server,
and be automatically informed about changes of these variables. Events can be
sent from SPEC to the clients.

The SPEC server does execute one macro at a time. It will queue macro
executions if the macro unit is currently occupied. This restriction
is the easiest way to avoid uncontrolled mixing of experimental
procedures. There are three points why this restriction is not very
important for practical purposes.
\begin{enumerate}
\item 
Not many incoming requests will lead to the execution of a macro. 
For example, it is possible to ask the value of a global variable while 
a scan macro is executing.

\item 
The event nature of the communication leads to a reduction in the client
server traffic. The client is automatically informed when something
interesting for him changes (i.e. a motor stops).

\item 
The level of commands. The client tells the server for example to
start a movement but will then be free to do other things. A graphical
user interface can start a movement on a motor, start later another 
movement of another motors and all that while the server is counting
on a detector.

\end{enumerate}

The SPEC server and client can be started and stopped independently and 
will reconnect automatically. There currently exists a Python client 
library and other SPEC version an be used as clients too. It is for
example possible to configure a motor in a client SPEC version to refer
to a motor in a SPEC server. One client can talk to multiple servers.

The development of the SPEC server mode has been accompanied by a new method
of writing pseudo motors and counters. Only these pseudo motors will be
fully functional in server mode. The new method consists of calling predefined
macro code from SPECs built in C code.

The server mode and pseudo motors are already pretty stable with only few
functionalities missing.

%___________________________________________________________________________

\hypertarget{the-plotting-framework}{}
\section*{The plotting framework}
\pdfbookmark[0]{The plotting framework}{the-plotting-framework}

PyDVT is a framework for visualization tools. It defines different objects 
and the data flow through them. The data flows from the "source" to the 
graphical "view" window. On its way it passes through "filters". There are
special selection classes to select which data to take from the source and
also selection classes which refer to the graphical selection done by the
user with the mouse.
\begin{description}
\item[view]

The views can be 2D display, contour plots, normal 1D plots, radial plots,
3D plot, data tables, or mesh plots. The underlying graphics kit is
PLPLOT to which we added user interaction. Online display of fast 2D
detectors requires a level of performance we could not get in this way.
We therefore used a specificly written library in C and a pixmap widget
for the 2D display.

\item[source]

For the moment the data can come from either files or SPEC shared memory.
The source also includes a trigger mechanism when the data has been updated.
The user of the tool kit does not have to be concerned with the update of
the plot in this way.

\item[filters]

It is relative easy to insert filters into the data-flow. There are 
filter fit-routines which take input from the data source and result in the
fitted curve (which can then be displayed in a view possibly together with
the raw data)

\end{description}

The toolkit is independent of the GUI toolkit. We currently use bindings for
Qt and Tk.

%___________________________________________________________________________

\hypertarget{visualization-applications}{}
\section*{Visualization Applications}
\pdfbookmark[0]{Visualization Applications}{visualization-applications}

General graphical applications have been created with the Python toolkit. 
Examples of these applications are:
\begin{description}
\item[PyMCA]

PyMCA is an application to work with spectra from multi channel analyzers.
It can read the spectra from SPEC files or SPEC shared memory and display
them. One can calibrate the spectra in user units (normally energy). Peak
search and fitting are integrated.

\item[PyDis]

PyDis takes two dimensional data from files or SPEC shared memory and
displays it in different ways. The application can be extended with plug-ins.
A poster presented at the workshop shows an example of PyDis extended
for data analysis ()

\item[NewPlot]

Displays online or offline SPEC data from scans. Allows fitting of the data 
and all the common view operations.

\end{description}
\includegraphics{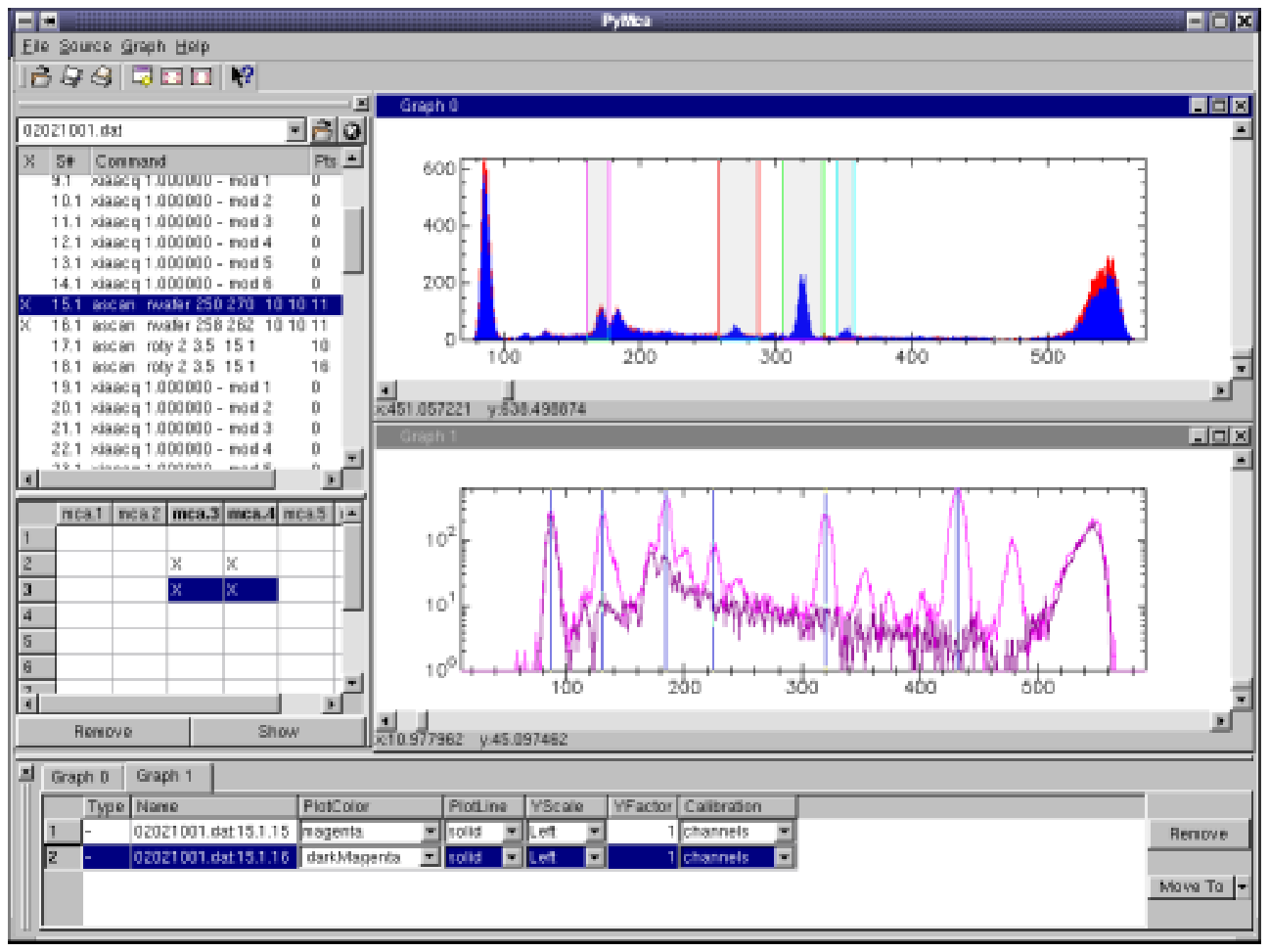}

The applications are finished but not still need some validating before we
can use them on our beamlines on a large scale.

%___________________________________________________________________________

\hypertarget{a-configuration-editor}{}
\section*{A Configuration Editor}
\pdfbookmark[0]{A Configuration Editor}{a-configuration-editor}

A common problem on our beamlines is the complexity of the configuration of
its components. To add a device to the system we often have to add information
in many different places. We decided therefore to create a configuration editor
which could be easily adapted to input all the necessary information. It
will replace the SPEC configuration editor and ESRF specific tools for 
configuration at our beamlines. It is also planned to use the editor in the 
NICOS control program of the Munich FRM II beamlines.

It consists of two parts:
\begin{enumerate}
\item 
A description editor which creates files which defines all the possible 
items to configure. It allows to create "properties" (i.e. the motor 
parameter "velocity", or the unit). Every property has a type and a 
default value. One can then define the properties of the objects in the 
system (i.e. that all motor objects have a "velocity" property).

\item 
A configuration editor which will use the information from the description
editor to create the above objects and configure their properties. The
editor allows to create groups (i.e. diffractometer or monochromator),
enable and disable parts of the configuration, and generally assists the
user during the configuration.

\end{enumerate}

The program is in the final development phase. We still need to provide 
output filters to save the final configuration in the standard ("old") 
file formats (i.e. SPEC config file).

%___________________________________________________________________________

\hypertarget{the-general-gui-framework}{}
\section*{The general GUI framework}
\pdfbookmark[0]{The general GUI framework}{the-general-gui-framework}

The general GUI framework is an attempt to integrate different standard
graphical components with beamline specific parts. Starting from an empty
application the individual elements are loaded interactively.  This
configuration is then saved. The general concept is born out of the experience
from our generic SPEC GUI and specific GUI applications (i.e. ProDC for the
protein crystallography beamlines). Both approaches had their advantages and
disadvantages. While the specific interface pleased which its rich
functionality we have to invest a lot of time to maintain it. We want to
combine complicated graphical standard components with beamline specific
panels.  The framework consists therefore of two parts:
\begin{enumerate}
\item 
A wizard to load in Python classes into the application. These classes 
need to be Qt or Tk Widgets (both at the same time don't work well at the 
moment). There is an optional mechanism to communicate with other classes
with an "eventhandler" (classes can send and receive events).

A special type of classes is the placer type class. These "placers" once
loaded into the applications accept other classes inside their window
in a predefined way (a up/down placer might accept two windows it will 
place either in the upper or lower half of its window).

The wizard will therefore go through the following steps:
\begin{quote}
\begin{itemize}
\item 
ask a filename and make the user select a Python class

\item 
ask the user where to put it on the screen (either in its own window
or inside another one)

\item 
ask additional parameters and create and instance

\end{itemize}
\end{quote}

\item 
An integrated editor for simple panels with simple graphical elements 
like input widgets, buttons, frames, labels, and so on. These panels will 
display information from SPEC and send commands to SPEC.

\end{enumerate}

This project is in its final stage but needs some more work.

%___________________________________________________________________________

\hypertarget{faq}{}
\section*{FAQ}
\pdfbookmark[0]{FAQ}{faq}
\begin{description}
\item[\emph{Why not build a scripting language on top of standard graphical components?}]

We need to provide fast (sometimes in hours) a 90{\%} solution. Our highest
priority is to make an experiment possible however inconvenient. We will
later try to automate certain steps and only later add a graphical 
interface for convenience.

\item[\emph{Why not build a system where Python talks directly to the device servers?}]

Apart from the obvious reason that SPEC had been more or less our only 
option when we started, there is a very import basic philosophy behind this 
decision - "Configure - don't program". SPEC allows us to hide all the 
different server implementations behind a standard layer. It allows us to 
adapt the system to the different experimental stations. Writing a specific 
application for an experimental station is relatively simple - but it is 
extremely difficult to maintain these type of programs. This approach 
becomes also inefficient with the large number of experimental stations as 
at the ESRF.

\end{description}

%___________________________________________________________________________

\hypertarget{conclusion}{}
\section*{Conclusion}
\pdfbookmark[0]{Conclusion}{conclusion}

Python has been proven an excellent tool. It allows to quickly develop
complex programs and helps to collaborate in the our group. On the
down side we see that the installation of the finished software takes
more effort and regret the absence of a powerful standard GUI toolkit. 
Our overall opinion stays however very positive.

The new concept is finished but some building blocks still need some more work.
News about our progress will be available on 
\href{http://www.esrf.fr/computing/bliss/bliss.html}{http://www.esrf.fr/computing/bliss/bliss.html} .

\end{document}